\documentclass[aps, reprint, prb, superscriptaddress]{revtex4-1}
\usepackage{graphicx}
\usepackage[caption=false]{subfig}    	
\usepackage{natbib}
\usepackage{multirow}
\usepackage{lipsum}
\usepackage{dcolumn}
\usepackage{float}
\usepackage{subfig}

\begin{document}
\title{Evidence for Superconducting Phase Coherence in the Metallic/Insulating Regime of the LaAlO$_3$--SrTiO$_3$ Interface Electron System}
\author{E. Fillis-Tsirakis}
\affiliation{Max Planck Institute for Solid State Research, 70569 Stuttgart, Germany}
\author{C. Richter}
\affiliation{Max Planck Institute for Solid State Research, 70569 Stuttgart, Germany}
\affiliation{Center for Electronic Correlations and Magnetism, Augsburg University, 86135 Augsburg, Germany}
\author{J. Mannhart}
\author{H. Boschker}
\email{h.boschker@fkf.mpg.de}
\affiliation{Max Planck Institute for Solid State Research, 70569 Stuttgart, Germany}


\begin{abstract}

A superconducting phase with an extremely low carrier density of the order of $\rm 10^{13}$ $\rm cm^{-2}$ is present at LaAlO$_3$--SrTiO$_3$ interfaces. If depleted from charge carriers by means of a gate field, this superconducting phase undergoes a transition into a metallic/insulating state that is still characterized by a gap in the spectral density of states. Measuring and analyzing the critical field of this gap, we provide evidence that macroscopically phase-coherent Cooper pairs are present in the metallic/insulating state. This is characterized by fluctuating vortex-antivortex pairs, and not by individual, immobile Cooper pairs. The measurements furthermore yield the carrier-density dependence of the superconducting coherence length of the two-dimensional system.

\end{abstract}

\maketitle

\section{Introduction}

Two-dimensional (2D) electron systems are fascinating, including phenomena such as the quantum Hall effect \cite{klitzing,tsukazaki}, ferromagnetism \cite{oja,vaz} and superconductivity \cite{goldman,gozar,reyren,ge,zhang}. The 2D-superconducting state yields substantial critical temperatures despite the presence of phase and amplitude fluctuations of the order parameter. The large susceptibility to fluctuations results in many cases in a bosonic 2D superconductor--to--insulator (SIT) transition \cite{shalnikov,haviland,valles,goldman,yazdani,fisher1,lee} or 2D superconductor--to--metal--to--insulator transition (SMIT) \cite{ephron,mason,seo,qin,humbert,li,park,phillips,jaeger,mason2,christiansen} with Cooper pairs existing in the insulating state \cite{fisher2}. These transitions are generally induced by tuning disorder or by changing the carrier density \cite{baturina,sacepe1,dubi,biscaras,stewart,barber,white,hollen,poran,sherman,noat,bollinger,shi,chand}. The LaAlO$_3$--SrTiO$_3$ interface 2D electron liquid (2DEL) \cite{ohtomo,breitschaft} is a 2D-superconductor that can be driven into an insulating state by depleting charge carriers with an electric field \cite{caviglia,schneider,mehta2,hurand,caprara}. This superconductor is of great interest because the superconducting state exists at extremely small carrier densities.

The transition between the superconducting state and the metallic/insulating state in the LaAlO$_3$--SrTiO$_3$ 2DEL has been well established by transport measurements \cite{caviglia,schneider,mehta2,caprara,lin,dikin,bell,richter,herranz}, however, with a different shape of the gate-voltage dependent $R(T)$ characteristics. Here $R$ is the sheet resistivity and $T$ is the temperature. Caviglia \textit{et al}. presented characteristics with d$R$/d$T$ $<$ 0 for gate voltages below a critical value and characteristics with d$R$/d$T$ $>$ 0 or a well defined zero resistivity state for gate voltages above this critical value. Intriguingly, at the critical value of the gate voltage, where d$R$/d$T$ = 0, the sheet resistivity matches the quantum resistance of paired electrons $h/4e^2$, ref.\cite{caviglia,lin}. This indicates a bosonic SIT, with Cooper pairs present in the insulating state. Other experiments, however, offer a less clear picture. One experiment observed an SIT at a resistance value equal to one third of $h/4e^2$, ref.\cite{dikin}, and others observed characteristics with d$R$/d$T$ = 0 for a large range of gate voltages \cite{hurand,bell,richter,herranz}.  In these experiments there is no single separatrix between the superconducting and insulating state, similar to the SMIT. We therefore refer to the non-superconducting phase as the metallic/insulating phase. These metallic phases have been argued to be bosonic in nature \cite{phillips}.

\begin{figure}[t]
\centering
\includegraphics[width=5.5cm]{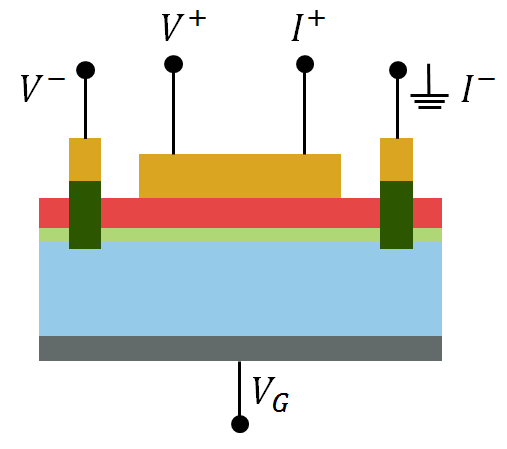}
\caption{The schematic of a tunnel junction on the measured sample. In detail: the four unit-cell thick LaAlO$_3$ layer is shown in red, the millimeter-thick SrTiO$_3$ layer in light blue, the 2DEL in light green, the gold layers in yellow, dark green is for the titanium contacts to the 2DEL and grey is the silver backgate electrode. The device current and the gate voltage were applied with respect to the ground contact of the 2DEL and the device voltage was measured between the top-electrode and another 2DEL contact.}
\label{fig1}
\end{figure}

Both scenarios suggest that Cooper pairs are present in the metallic/insulating state of the LaAlO$_3$--SrTiO$_3$ 2DEL. Furthermore, tunneling measurements find a superconducting gap in the density of states (DOS) across the transition \cite{richter}. Depending on their interaction, the Cooper pairs possibly form a macroscopic quantum-state characterized by an order parameter, a coherence length, and a critical magnetic field. It is also possible that the Cooper pairs act as single particles if their interaction is non-existent or weak. Which of the two scenarios occurs in a superconductor with small carrier density is unknown, yet important for the general understanding of superconductivity. The two scenarios can be fundamentally different, as exemplified by their response to applied (perpendicular) magnetic fields $H$. If the Cooper pairs are phase coherent, the perpendicular upper critical field $H_{\rm c}$ is determined by vortex behavior, whereas if the Cooper pairs are localized without phase coherence, $H_{\rm c}$ is determined by pair breaking due to the Zeeman energy. In the above cases, $H_{\rm c}$ differs considerably and for LaAlO$_3$--SrTiO$_3$ it equals 0.3 T and 0.8 T, respectively, as discussed below. This difference opens a route to determine unequivocally the Cooper-pair nature of the metallic/insulating state, as $H_{\rm c}$ can be well measured. To measure $H_{\rm c}$ precisely, we use magnetic-field-dependent tunneling spectroscopy. Resistivity measurements also allow us to determine $H_{\rm c}$, but are less stringent for interpreting in the metallic/insulating state. In tunneling spectroscopy measurements, the disappearance of the superconducting gap in the spectra quantitatively yields $H_{\rm c}$ for the superconducting as well as for the metallic/insulating sides of the SMIT. As described in this letter, we observe $H_{\rm c}$ values across the SMIT that are in clear agreement with the vortex--physics scenario. The data therefore provide conclusive evidence that the Cooper pairs in the metallic/insulating side are phase coherent on a length scale of at least the vortex size.

\begin{figure}[t]
\centering
\includegraphics[width=12cm]{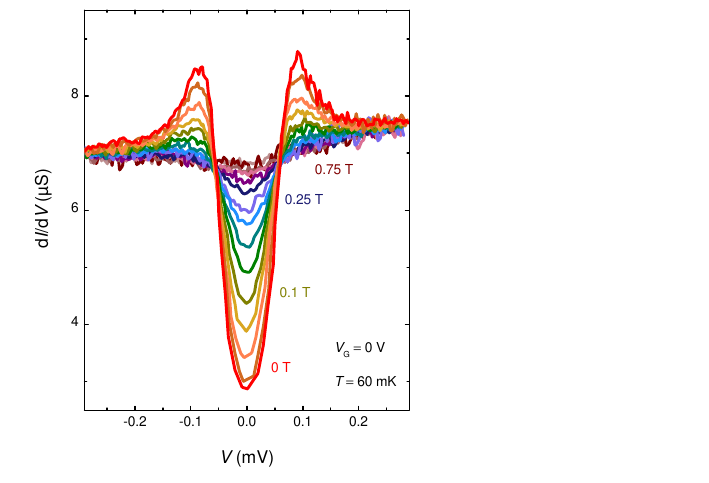}
\caption{The $H$ dependence of the d$\it{I}$$/$d$\it{V}$($\it{V}$) characteristics at zero back-gate field. Tunnel spectra were obtained at several values of $H$, with an incremental step of 0.025 T for $H < 0.2$ T; the larger values are 0.25, 0.3, 0.4, 0.5, 0.75 and 1 T.}
\label{fig1}
\end{figure}

\begin{figure*}[t]  
    \mbox{\includegraphics[width=5.46cm]{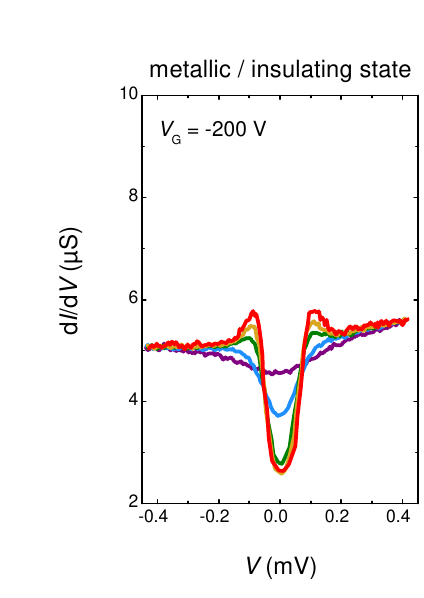}}   
    \hspace{3px}
    \mbox{\includegraphics[width=3.8cm]{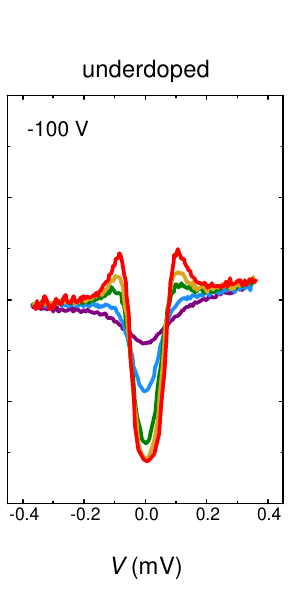}}
    \hspace{0px}
    \mbox{\includegraphics[width=3.8cm]{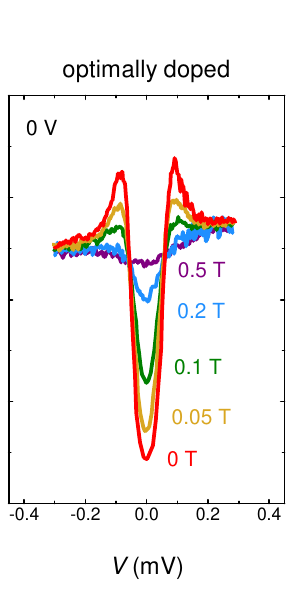}}
    \hspace{0px}
    \mbox{\includegraphics[width=3.8cm]{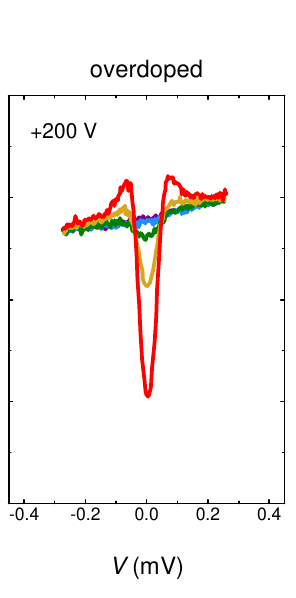}}
    \hspace{8px}
    \caption{Gate-voltage dependence of the tunnel spectra measured at 60 mK, as a function of magnetic field $H$. The superconducting gap for gate voltages of $-$200, $-$100, 0, 200~V is suppressed at $H$ = 300$\pm$50, 300$\pm$50, 215$\pm$30  and  90$\pm$20 mT, respectively. The error margins were derived by analyzing a series of spectra spaced from 20 to 50 mT (not shown here). In the metallic / insulating state the superconducting gap and the coherence peaks are visible even at gate voltages for which no resistive superconducting transition is observed.}
    \label{fig2}
\end{figure*}

\section{Experimental}

To perform the tunneling spectroscopy measurements, planar junctions were realized, with device structures that are discussed in detail elsewhere \cite{richter}. A schematic of the junction structure is shown in Fig. 1. A Au topelectrode is deposited on top of the four unit cell thick LaAlO$_3$ layer that simultaneously generates the 2DEL and acts as a tunnel barrier. Different Ti contacts to the 2DEL allow for four-point measurements of the tunneling current. The device area is approximately 1 $\rm mm^2$. Tunnel spectra were obtained by applying a current between the top electrode and the interface and by simultaneously recording the tunneling voltage as well as the $ac$ conductance using a standard lock-in technique. The polarity of the voltage characterizes the sign of the interface voltage with respect to the top electrode bias; for $V<0$ electrons tunnel out of the 2DEL. Measurements with these junctions resolved the superconducting gap of the 2D-state. The carrier concentration of the 2DEL was tuned electrically by a back-gate voltage $V_{\rm G}$, allowing tunnel measurements across the entire superconducting dome (maximum $T_{\rm c}$ $\approx$ 300 mK) as well as in the metallic/insulating state. Even in the insulating state, the minimal tunneling resistance significantly exceeds the maximal resistance of the 2DEL. All measurements were done at a temperature of $\sim$60 mK. The $\rm {SrTiO_3}$ substrate had previously undergone an isotope exchange, with $^{18}$O substituting for $^{16}$O. We did not see an effect of the exchange on the transport properties of the 2DEL\cite{richter,boschker}.

\section{Results}

To measure $H_{\rm c}$, the disappearance of the superconducting gap was analyzed as a function of magnetic field $H$. Figure~\ref{fig1} shows tunnel spectra as a function of $H$, with $V_{\rm G}$ = 0. In these tunnel junctions, the differential conductance reflects the DOS of the 2DEL, and the superconducting gap $\it \Delta$ is observed with a value of $\sim$60~$\rm~\mu$V. The suppression of the density of states at $V \rm = 0$ and the quasiparticle peaks disappear gradually with increasing $H$, as expected for type-II superconductors. The measured tunnel spectra represent a spatial average of the superconducting gap because the tunnel junctions are much larger than the vortex size. Therefore, for $\it H \ll H_{\rm c}$, the tunnel spectra are magnetic-field independent as the volume fraction of the vortices is negligible. For larger values of $H$ that are still smaller than $H_{\rm c}$, the measured spectra have reduced and broadened coherence peaks and a large conductance at $V = 0$, owing to a reduced superconducting gap across a significant fraction of the device area. Close to $H_{\rm c}$, the vortices overlap and the maximum gap is reduced. Finally, for $H > H_{\rm c}$, the spectra are weakly dependent on the magnetic field and the conductance has only a small reduction when $V\rightarrow 0$. This reduction is commonly attributed to the Altshuler--Aronov correction \cite{altshuler} that accounts for electron--electron interactions. The Altshuler--Aronov correction can be easily distinguished from the superconducting gap, because it has a different energy scale (it persists up to approximately 1 meV) and a diffferent $T$ and $H$ dependence. 

We now turn to the carrier-density dependence of the critical field. The d$\it{I}$$/$d$\it{V}$($\it{V}$) spectra for four different values of gate voltage are shown in Fig.~\ref{fig2}. The four panels cover the entire density range, from the overdoped to the metallic/insulating side, as shown by the $R$($T$) characteristics of the 2DEL. In all cases the superconducting gap is present at 0 T. The gap increases with decreasing $V_{\rm G}$, $i.e.$, decreasing carrier concentration $n$. In addition, with the decrease of $n$, the normal-state conductance is also reduced. Even in the metallic/insulating state, clear quasiparticle coherence peaks are observed in the spectra. The applied magnetic field increases the conductivity at $V = 0$ and reduces the quasiparticle peaks. The superconducting gap features are no longer observed at fields larger than 0.3 T for all values of the gate voltage. The doping level also controls the d$\it{I}$$/$d$\it{V}$($V \rm{=0},\it{H}$) behavior. With decreasing carrier density, larger magnetic fields are required to increase the d$\it{I}$$/$d$\it{V}$($V \rm{=0}$) conductance towards the normal-state level.

\begin{figure}[ht]
\centering
\includegraphics[width=9cm]{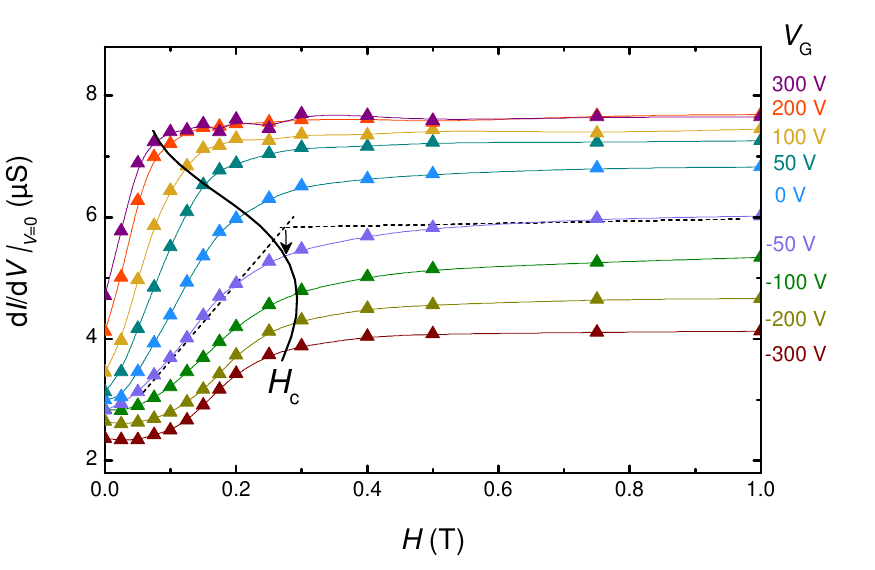}
\caption{The d$\it{I}$$/$d$\it{V}$($V \rm{=0},\it{H}$) characteristics for $-300 < V_{\rm g} < 300$ V. $H_{\rm c}$ is derived as illustrated.}
\label{fig3}
\end{figure}

\begin{figure}[h]
\begin{minipage}{\columnwidth}
\includegraphics[width=9.7cm]{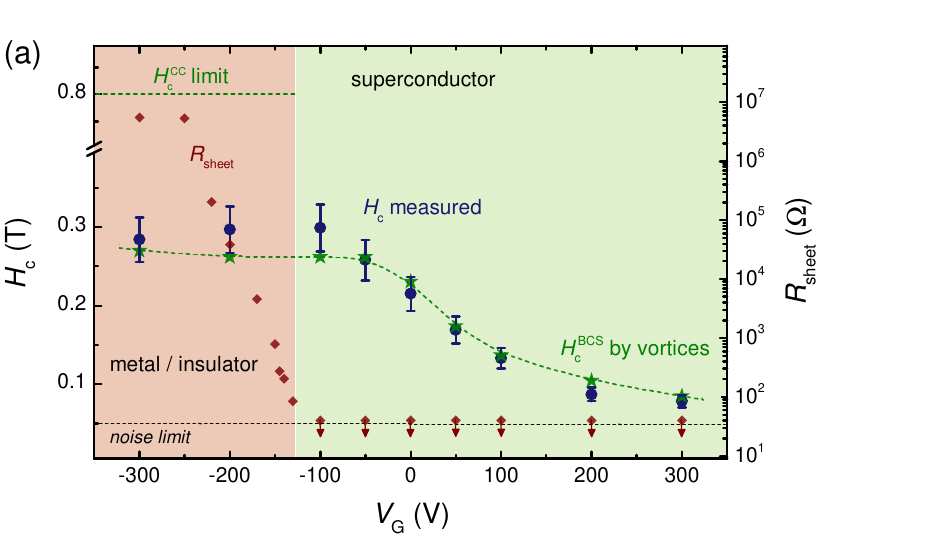}
\end{minipage}
\begin{minipage}{\columnwidth}
\includegraphics[width=8.45cm]{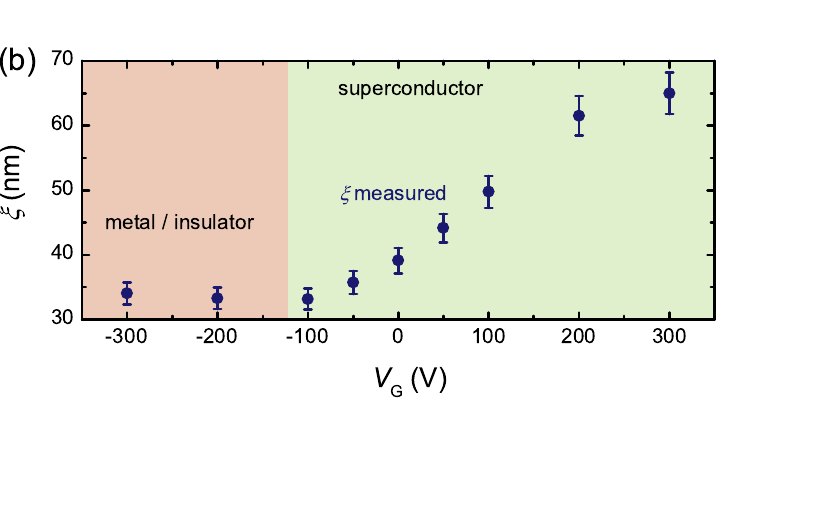}
\end{minipage}
\caption{a) Comparison between the measured $H_{\rm c}$ and predicted values by the individual Cooper pair ($H_{\rm c}^{\rm CC}$ limit) and vortex ($H_{\rm c}^{\rm BCS}$ by vortices) models. The measured data agrees well with the vortex prediction model. The colored background separates the superconductor and metal / insulator regimes at $V_{\rm g} \approx -130$ V, as shown by the resistivity data. The error margins are derived from the extrapolation procedure shown in Fig. 4. b) The gate-voltage dependence of the in-plane coherence length $\xi$. }
\label{fig4}
\end{figure}

Fig.~\ref{fig3} shows the d$\it{I}$$/$d$\it{V}$($V \rm{=0}$) as a function of $H$, in order to illustrate the transition to the normal state more precisely. For all gate voltages, d$\it{I}$$/$d$\it{V}$($V \rm{=0}$) increases until a plateau is reached. The plateau reflects the absence of the superconducting gap. By linearly extrapolating the curves at the steepest point of d$\it{I}$$/$d$\it{V}$($V \rm{=0},\it{H}$) we determine $H_{\rm c}$, as shown in Fig.~\ref{fig3}. We now discuss the carrier density dependence of $\it H_{\rm c}$. In Fig.~\ref{fig4}a the measured values for $\it H_{\rm c}$ are given as a function of the backgate voltage. The low temperature resistivity, at $H$ = 0, of the 2DEL is shown as well, indicating the SMIT. $H_{\rm c}$ monotonically increases with decreasing charge carrier density. Starting at 80 mT in the overdoped range, $H_{\rm c}(V_{\rm G})$ reaches 300 mT in the underdoped range. Also in the metallic/insulating regime  ($-200$ V and $-300$ V), $H_{\rm c}$ $\approx$ 300 mT. 

\section{Discussion}

Now we compare the critical fields with the Chandrasekhar--Clogston critical field \cite{Chandrasekhar, Clogston} and the one induced by vortices. For the latter case, $H_{\rm c}$ ($\equiv$ $H_{\rm c2}$) is obtained from the Ginzburg Landau coherence length $\xi$ by $H_{\rm c2}=\Phi_{\rm 0}/2\pi\xi^2$, where $\Phi_{\rm 0} = h/2e$. In the BCS model, $\xi$ is related to the superconducting gap by $\it \xi = \hbar v_{\rm f}/\pi \Delta$. Here, $v_{\rm f}$ is the Fermi velocity. The Ginzburg Landau coherence length is equal to the BCS coherence length for superconductors that are in the clean limit. In the case of the LaAlO$_3$--SrTiO$_3$ superconductor, the coherence length is of similar magnitude to the transport scattering length (20 to 100 nm \cite{huijben,dagan,mannhart}) (see below). We therefore expect the BCS coherence length to be similar to the Ginzburg Landau coherence length. To calculate $\xi$, we determine $\it \Delta(V_{\rm G}) $ from the tunnel spectra at zero applied magnetic field. It ranges from 37 to 66 $\mu$eV. Equating $H_{\rm c}^{\rm BCS}$ to the measured $H_{\rm c}$ in the superconducting region ($-$100 V $\leq$ $V_{\rm G}$ $\leq$ 300 V), we determine $v_{\rm f}$ to be $\rm 1.1 \times 10^4$ $\rm m/s$, thus it is found to be independent of the gate voltage. This value is smaller than the theoretically predicted $\rm 7 \times 10^4-5 \times 10^5$ $\rm m/s$ \cite{nakamura}. We note that the chemical potential changes only by a small amount in this gate voltage range \cite{boschker}, thus $v_{\rm f}$ is also expected to show only small changes. The good agreement between $H_{\rm c}^{\rm BCS}$ and measured $H_{\rm c}$ also holds in the metallic/insulating region (assuming $v_{\rm f}$ stays also constant in this region), and the entire gate voltage dependence of $H_{\rm c}$ can be well described by the model based on the existence of BCS-type vortices. The increase of $H_{\rm c}$ with decreasing carrier density is therefore due to the increasing size of the superconducting gap. In contrast, the predictions by the Chandrasekhar-Clogston limit ($H_{\rm c}^{\rm CC}$) are in disagreement with the measured data. $H_{\rm c}^{\rm CC}$ is given by $\it \Delta \rm / \sqrt{2} \mu_{\rm B}$, where $\mu_{\rm B}$ is the Bohr magneton. Here, the standard $g$-factor of 2 is assumed. With $\it \Delta$ = 66 $\mu$eV (the constant value of the gap in the metallic/insulating region) this ratio yields $H_{\rm c}^{\rm CC}$ = 0.8 T. This value does not include the spin-orbit coupling \cite{Cavigliax, Kim} and is therefore a lower-limit estimate of the upper critical field due to depairing.

The previous analysis yields an accurate determination of the in-plane coherence length $\xi$ as a function of gate voltage. It ranges from 34 to 65 nm across the phase diagram, Fig. 5b. These values for $\xi$ are lower than those reported in previously published work \cite{reyren2,gariglio,benshalom,dikin,mehta,herranz}, which are typically around 70 nm, increasing with increasing carrier density \cite{benshalom}. The main reason is due to the difference of the methods used to determine $H_{\rm c2}$. In the present work, the disappearance of the superconducting gap is used as a criterion to identify $H_{\rm c2}$. In earlier work, $H_{\rm c2}$ has been determined as the magnetic field at which the sample resistance equals 50 \% of the normal state resistance. In electron systems with a broad superconducting transition, the latter criterion underestimates $H_{\rm c2}$, resulting in correspondingly larger values of the coherence length. 

\section{Conclusion}
The measured critical field $H_{\rm c} \approx$ 0.28 T is consistent only with the value expected for the suppression of a vortex phase; the localized Cooper pair scenario does not match the measured data. Because the coherence length $\xi$ also evolves gradually from the macroscopically phase-coherent, superconducting state to the metallic/insulating state, the measurements provide evidence that the electron system in the metallic/insulating state consists of one or more ensembles of macroscopically phase-coherent Cooper pairs. This understanding is corroborated by the existence of the coherence peaks in the tunnel spectra in the metallic/insulating state \cite{sacepe2}. The data exclude an electron system consisting solely of superconducting puddles with a length scale of $l < \xi$.

\section{acknowledgement}
We thank M. Etzkorn for helpful discussions.


\begin {thebibliography}{3}

\bibitem{klitzing} K. v. Klitzing, G. Dorda and M. Pepper, Phys. Rev. Lett. $\bf 45$, 494 (1980).
\bibitem{tsukazaki} A. Tsukazaki, A. Ohtomo, T. Kita, Y. Ohno, H. Ohno and M. Kawasaki, Science $\bf 385$, 1388 (2007).
\bibitem{oja} R. Oja, M. Tyunina, L. Yao, T. Pinomaa, T. Kocourek, A. Dejneka, O. Stupakov, M. Jelinek, V. Trepakov, S. van Dijken and R. M. Nieminen, Phys. Rev. Lett. $\bf 109$, 127207 (2012).
\bibitem{vaz} C. A. F. Vaz, J. A. C. Bland and G. Lauhoff, Reg. Prog. Phys. $\bf 71$, 056051 (2008).
\bibitem{gozar} A. Gozar, G. Logvenov, L. Fitting Kourkoutis, A. T. Bollinger, L. A. Gianuzzi, D. A. Muller and I. Bozovic, Nature $\bf 455$, 782 (2008).
\bibitem{reyren} N. Reyren, S. Thiel, A. D. Caviglia, L. Fitting Kourkoutis, G. Hammerl, C. Richter, C. W. Schneider, T. Kopp, A.-S. Ruetschi, D. Jaccard, M. Gabay, D. A. Muller, J.-M. Triscone and J. Mannhart, Science $\bf 317$, 1196 (2007).
\bibitem{ge} J.-F. Ge, Z.-L. Liu, C. Liu, C.-L Gao, D. Qian, Q.-K. Xue, Y. Liu and J.-F Jia, Nature Mater. $\bf 14$, 285 (2014).
\bibitem{zhang} W. H. Zhang, Y. Sun, J.-S. Zhang, F.-S. Li, M.-H. Guo, Y.-F. Zhao, H.-M. Zhang, J.-P. Peng, Y. Xing, H.-C. Wang, T. Fujita, A. Hirata. Z. Li, H. Ding, C.-J. Tang, M. Wang, Q.-Y. Wang, K. He, S.-H. Ji, X. Chen, J.-F. Wang, Z.-C. Xia, L. Li, Y.-Y. Wang, J. Wang, L.-L. Wang, M.-W. Chen, Q.-K. Xue and X.-C. Ma, Chin. Phys. Lett. $\bf 31$, 017401 (2014).
\bibitem{goldman} A. M. Goldman and N. Markovic, Phys. Today $\bf 51$, 39 (1998).
\bibitem{shalnikov} A. I. Shalnikov, Nature $\bf 142$, 74 (1938).
\bibitem{haviland} D. B. Haviland, Y. Liu and A. M. Goldman, Phys. Rev. Lett. $\bf 62$, 2180 (1989).
\bibitem{valles} J. M. Valles, R. C. Dynes and J. P. Garno, Phys. Rev. Lett. $\bf 69$, 3567 (1992).
\bibitem{yazdani} A. Yazdani and A. Kapitulnik, Phys. Rev. Lett. $\bf 74$, 3037 (1995).
\bibitem{fisher1} M. P. A. Fisher, G. Grinstein and S. M. Girvin, Phys. Rev. Lett. $\bf 64$, 587 (1990).
\bibitem{lee} P. A. Lee and T. V. Ramakrishnan, Rev. Mod. Phys. $\bf 57$, 287 (1985).

\bibitem{ephron} D. Ephron, A. Yazdani, A. Kapitulnik and M. R. Beasley, Phys. Rev. Lett. $\bf 76$, 1529 (1996).
\bibitem{mason} N. Mason and A. Kapitulnik, Phys. Rev. Lett. $\bf 82$, 5341 (1999).
\bibitem{seo} Y. Seo, Y. Qin, C. L. Vicente, K. S. Choi and J. Yoon, Phys. Rev. Lett. $\bf 97$, 057005 (2006).
\bibitem{qin} Y. Qin, C. L. Vicente and J. Yoon, Phys. Rev. B $\bf 73$, 100505(R) (2006).
\bibitem{humbert} V. Humbert, F. Couedo, O. Crauste, L. Berge, A.-A. Drillien, C. A. Marrache-Kikuchi and L. Dumoulin, J. Phys. Conf. Series $\bf 568$ 052012 (2014).
\bibitem{li} Y. Li, C. L. Vicente and J. Yoon, Phys. Rev. B $\bf 81$, 020505(R) (2010).
\bibitem{park} S. Park and E. Kim, arXiv$:$1401.3947 (2014).
\bibitem{phillips} P. Phillips and D. Dalidovich, Science $\bf 302$, 243 (2003).
\bibitem{jaeger} H. M. Jaeger, D. B. Haviland, A. M. Goldman and B. G. Orr, Phys. Rev. B $\bf 34$, 7 (1986).
\bibitem{mason2} N. Mason and A. Kapitulnik, Phys. Rev. B $\bf 64$, 060504(R) (2001).
\bibitem{christiansen} C. Christiansen, L. M. Hernandez and A. M. Goldman, Phys. Rev. Lett $\bf 88$, 3 (2002).

\bibitem{fisher2} M. P. A. Fisher, P. B. Weichman, G. Grinstein and D. S. Fisher, Phys. Rev. B $\bf 40$, 546 (1989).
\bibitem{baturina} T. I. Baturina, A. Y. Mironov, V. M. Vinokur, M. R. Baklanov and C. Strunk, Phys. Rev. Lett. $\bf 99$, 257003 (2007).
\bibitem{sacepe1} B. Sacepe, C. Chapelier, T. I. Baturina, V. M. Vinokur, M. R. Baklanov and M. Sanquer, Nature Commun. $\bf 1$, 140 (2010).
\bibitem{dubi} Y. Dubi, Y. Meir and Y. Avishai, Nature $\bf 449$, 876 (2007).
\bibitem{biscaras} J. Biscaras, N. Bergeal, S. Hurand, C. Feuillet-Palma, A. Rastogi, R. C. Budhani, M. Grilli, S. Caprara and J. Lesqueur, Nature Mater. $\bf 12$, 542 (2013).
\bibitem{stewart} M. D. Stewart Jr., A. Yin, J. M. Xu, and J. M. Valles Jr., Science $\bf 318$, 1273 (2007).
\bibitem{barber} R. P. Barber Jr., L. M. Merchant, A. La Porta and R. C. Dynes, Phys. Rev. B $\bf 49$, 3409 (1994).
\bibitem{white} A. E. White, R. C. Dynes and J. P. Garno, Phys. Rev. B $\bf 33$, 3549 (1986).
\bibitem{hollen} S. M. Hollen, H. Q. Nguyen, E. Rudisaile, M. D. Stewart Jr., J. Shainline, J. M. Xu and J. M. Valles Jr., Phys. Rev. B $\bf 84$, 064528 (2011).
\bibitem{poran} S. Poran, E. Shimshoni and A. Frydman, Phys. Rev. B $\bf 84$, 014529 (2011).
\bibitem{sherman} D. Sherman, G. Kopnov, D. Shahar and A. Frydman, Phys. Rev. Lett. $\bf 108$, 177006 (2012).
\bibitem{noat} Y. Noat, V. Cherkez, C. Brun, T. Cren, C. Carbillet, F. Debontridder, K. Ilin, M. Siegel, A. Semenov, H.-W. Hubers and D. Roditchev, Phys. Rev. B. $\bf 88$, 014503 (2013).
\bibitem{bollinger} A. T. Bollinger, G. Dubuis, J. Yoon, D. Pavuna, J. Misewich and I. Bozovic, Nature $\bf 472$, 458 (2011).
\bibitem{shi} X. Shi, G. Logvenov, A. T. Bollinger, I. Bozovic, C. Panagopoulos and D. Popovic, Nature Mater. $\bf 12$, 47 (2012).
\bibitem{chand} M. Chand, G. Saraswat, A. Kamlapure, M. Mondal, S. Kumar, J. Jesudasan, V. Bagwe, L. Benfatto, V. Tripathi and P. Raychaudhuri, Phys. Rev. B $\bf 85$, 014508 (2012).
\bibitem{ohtomo} A. Ohtomo and H. Y. Hwang, Nature $\bf 427$, 423 (2004).
\bibitem{breitschaft} M. Breitschaft, V. Tinkl, N. Pavlenko, S. Paetel, C. Richter, J. R. Kirtley, Y. C. Liao, G. Hammerl, V. Eyert, T. Kopp and J. Mannhart, Phys. Rev. B $\bf 81$, 153414 (2010).
\bibitem{hurand} S. Hurand, A. Jouan, C. Feuillet-Palma, G. Singh, J. Biscaras, E. Lesne, N. Reyren, A. Barthelemy, M. Bibes, C. Ulysse, X. Lafosse, M. Pannetier-Lecoeur, S. Caprara, M. Grilli, J. Lesueur and N. Bergeal, Sci. Rep. $\bf 5$, 12571 (2015).
\bibitem{caviglia} A. D. Caviglia, S. Gariglio, N. Reyren, D. Jaccard, T. Schneider, M. Gabay, S. Thiel, G. Hammerl, J. Mannhart and J.-M. Triscone, Nature $\bf 456$, 624 (2008).

\bibitem{schneider} T. Schneider, A. D. Caviglia, S. Gariglio, N. Reyren and J.-M. Triscone, Phys. Rev. B $\bf 79$, 184502 (2009).
\bibitem{mehta2} M. M. Mehta, D. A. Dikin, C. W. Bark, S. Ryu, C. M. Folkman, C. B. Eom and V. Chandrasekhar, Phys. Rev. B $\bf 90$, 100506(R) (2014).
\bibitem{caprara} S. Caprara, D. Bucheli, N. Scopigno, J. Biscaras, S. Hurand, J. Lesueur and M. Grilli, Supercond. Sci. Technol. $\bf 28$, 014002 (2015).
\bibitem{lin} Y.-H. Lin, J. Nelson and A. M. Goldman, Physica C $\bf 514$, 130 (2015).
\bibitem{dikin} D. A. Dikin, M. Mehta, C. W. Bark, C. M. Folkman, C. B. Eom and V. Chandrasekhar, Phys. Rev. Lett. $\bf 107$, 056802 (2011).
\bibitem{richter} C. Richter, H. Boschker, W. Dietsche, E. Fillis-Tsirakis, R. Jany, F. Loder, L. F. Kourkoutis, D. A. Muller, J. R. Kirtley, C. W. Schneider and J. Mannhart, Nature $\bf 502$, 528 (2013).
\bibitem{herranz} G. Herranz, G. Singh, N. Bergeal, A. Jouan, J. Lesqueur, J. Gasquez, M. Varela, M. Scigaj, N. Dix, F. Sanchez and J. Fontcuberta, Nature Commun. $\bf 6$, 6028 (2015).
\bibitem{bell} C. Bell, S. Harashima, Y. Kozuka, M. Kim, B. G. Kim, Y. Hikita and H. Y. Hwang, Phys. Rev. Lett. $\bf 103$, 226802 (2009).



\bibitem{boschker} H. Boschker, C. Richter, E. Fillis-Tsirakis, C. W. Schneider and J. Mannhart, Sci. Rep. $\bf 5$, 12309 (2015).
\bibitem{altshuler} B. Altshuler and A. Aronov, Solid State Commun. $\bf 30$, 115 (1971).
\bibitem{Chandrasekhar} B. S. Chandrasekhar, Appl. Phys. Lett. $\bf 1$, 7 (1962).
\bibitem{Clogston} A. M. Clogston, Phys. Rev. Lett. $\bf 9$, 266 (1962).
\bibitem{mannhart} J. Mannhart and D. G. Schlom, Science $\bf 327$, 1607 (2010).
\bibitem{dagan} M. Ben Shalom, C. W. Tai, Y. Lereah, M. Sachs, E. Levy, D. Rakhmilevich, A. Palevski and Y. Dagan, Phys. Rev. B $\bf 80$, 140403(R) (2009).
\bibitem{huijben} M. Huijben, G. Rijnders, D. H. A. Blank, S. Bals, S. van Aert, J. Verbeeck, G. van Tendeloo, A. Brinkman and H. Hilgenkamp, Nature Mater. $\bf 5$, 556 (2006).
\bibitem{reyren2} N. Reyren, S. Gariglio, A. D. Caviglia, D. Jaccard, T. Schneider and J.-M. Triscone, Appl. Phys. Lett. $\bf 94$, 112506 (2009).
\bibitem{gariglio} S. Gariglio, N. Reyren, A. D. Caviglia and J.-M. Triscone, J. Phys. Condens. Matter $\bf 21$, 164213 (2009).
\bibitem{benshalom} M. Ben Shalom, M. Sachs, D. Rakhmilevitch, A. Palevski and Y. Dagan, Phys. Rev. Lett. $\bf 104$, 126802 (2010).
\bibitem{mehta} M. M. Mehta, D. A. Dikin, C. W. Bark, S. Ryu, C. M. Folkman, C. B. Eom and V. Chandrasekhar, Nature Commun. $\bf 3$, 955 (2012).
\bibitem{nakamura} Y. Nakamura and Y. Yanase, J. Phys. Soc. Jpn. $\bf 82$, 083705 (2013).
\bibitem{Cavigliax} A. D. Caviglia, M. Gabay, S. Gariglio, N. Reyren, C. Cancellieri and J.-M. Triscone, Phys. Rev. Lett. $\bf 104$, 126803 (2010).
\bibitem{Kim} M. Kim, Y. Kozuka, C. Bell, Y. Hikita and H. Y. Hwang, Phys. Rev. B $\bf 86$, 085121 (2012).
\bibitem{sacepe2} B. Sacepe, T. Dubouchet, C. Chapelier, M. Sanquer, M. Ovadia, D. Shahar, M. Feigelman and L. Ioffe, Nature Phys. $\bf 7$, 239 (2011).

\end{thebibliography}
\end{document}